\documentstyle[aps,twocolumn,floats,prl,psfig]{revtex}       

\newcommand{\ApJL}{Astrophys. J. Lett.}

\newcommand{\PRL}{Phys. Rev. Lett.}
\newcommand{\PRD}{Phys. Rev. D}

\newcommand{\vecl}{{\bf l}}    
\newcommand{\rad}{r} 
\newcommand{\da}{d_A}
\newcommand{\bn}{{\bf \hat{n}}}

\begin{document}
\twocolumn[\hsize\textwidth\columnwidth\hsize\csname
@twocolumnfalse\endcsname

\title{Separation of Gravitational-Wave and Cosmic-Shear
Contributions to Cosmic Microwave Background Polarization}
\author{Michael Kesden, Asantha Cooray, and Marc Kamionkowski}
\address{California Institute of Technology, Mail Code 130-33,
Pasadena, CA 91125\\ E-mail:
kesden@caltech.edu,asante@caltech.edu,kamion@tapir.caltech.edu}

\date{\today}

\maketitle

\begin{abstract}
Inflationary gravitational waves (GW) contribute to
the curl component in the polarization of the CMB.
Cosmic shear---gravitational lensing of the CMB---
converts a fraction of the dominant gradient
polarization to the curl component.  Higher-order
correlations can be used to map the cosmic shear
and subtract this contribution to the curl.
Arcminute resolution will be required to pursue GW
amplitudes smaller than those accessible by Planck.
The finite cutoff in CMB power at small scales
leads to a minimum detectable GW amplitude
corresponding to an inflation energy near $10^{15}$ GeV.
\end{abstract}
\vskip 0.5truecm

\pacs{98.80.Es,95.85.Nv,98.35.Ce,98.70.Vc
\hfill}
]

Observation of acoustic oscillations in the temperature
anisotropies of the cosmic microwave background (CMB;
\cite{Miletal99}) strongly suggests an inflationary origin for
primordial perturbations \cite{Gut81}. It has been argued that
a new smoking-gun
signature for inflation would be detection of stochastic
background of inflationary gravitational waves (IGWs)
\cite{KamKos99}.  These IGWs produce a distinct signature in the
CMB in the form of a contribution to the curl, or
magnetic-like, component of the polarization \cite{KamKosSte97}.
Since there is no scalar, or density-perturbation, contribution to
these curl modes, curl polarization  was considered to be a direct
probe of IGWs.  

There is, however, another source of a curl component.  Cosmic
shear (CS)---weak gravitational lensing of the CMB due to
large-scale structure along the line of sight---results in a
fractional conversion of the gradient mode from density
perturbations to the curl component \cite{ZalSel98}.  
The amplitude of the IGW background varies
quadratically with the energy scale $E_{\rm infl}$ of inflation,
and so the prospects for detection also depend on this
energy scale. In the absence of CS, the smallest
detectable IGW background scales simply with the sensitivity of
the CMB experiment---as the instrumental sensitivity is
improved, smaller values of $E_{\rm infl}$ become accessible
\cite{KamKos99,JafKamWan00}.  More realistically, however, the CS-induced
curl introduces a noise from which IGWs must be distinguished.
If the IGW amplitude (or $E_{\rm infl}$) is sufficiently large,
the CS-induced curl will be no problem.  However, as $E_{\rm
infl}$ is reduced, the IGW signal becomes smaller and will get
lost in the CS-induced noise.  This confusion leads to a minimum
detectable IGW amplitude \cite{LewChaTur02}.

In addition to producing a curl component, CS also introduces
distinct higher-order correlations in the CMB temperature
pattern.  Roughly speaking, lensing can stretch the image of the
CMB on a small patch of sky and thus lead to something akin to
anisotropic correlations on that patch of sky, even though the
CMB pattern at the surface of last scatter had isotropic
correlations.  By mapping these effects, the CS can be
mapped as a function of position on the sky \cite{SelZal99}.
The observed CMB polarization can then be corrected for these
lensing deflections to reconstruct the intrinsic CMB
polarization at the surface of last scatter (in which the only
curl component would be that due to IGWs).
In this {\it Letter} we evaluate how well this subtraction can
be accomplished and study the impact of CS on experimental
strategies for detection of IGWs.

To begin, we review the determination of the smallest detectable
IGW amplitude in the absence of CS.  Following
Ref. \cite{JafKamWan00}, we consider a CMB-polarization experiment
of some given instrumental sensitivity quantified by the
noise-equivalent temperature (NET) $s$, angular resolution
$\theta_{\rm FWHM}$, duration $t_{\rm yr}$ in years, and
fraction of sky covered $f_{\rm sky}$.  We then make the null
hypothesis of no IGWs and determine the largest IGW amplitude
${\cal T}$, defined as ${\cal T}=9.2 V/m_{\rm Pl}^4$, where
$V=E_{\rm infl}^4$ is the inflaton-potential height, that would
be consistent at the $1\sigma$ level with the null detection.
We then obtain the smallest detectable IGW amplitude
$\sigma_{\cal T}$ from
\begin{equation}
     \sigma_{\mathcal{T}}^{-2} = \sum_{l>180/\theta}
     \left( \partial
     C_{l}^{BB,GW} /\partial \mathcal{T} \right)^2 (\sigma_{l}^{BB})^{-2},
\label{eqn:sigmaT}
\end{equation}
where $C_l^{BB,GW}$ is the IGW contribution to the curl power
spectrum, and
\begin{equation}
\label{E:siggrav}
     \sigma_{l}^{BB} = \sqrt{\frac{2}{f_{sky}(2l+1)}} \left(C_{l}^{BB} +
     f_{sky} w^{-1} e^{l^2 \sigma_{b}^2} \right) \, ,
\label{eqn:sigma}
\end{equation}
is the standard error with which each multipole moment
$C_l^{BB}$ can be determined.  Here, $w^{-1} = 4\pi (s/T_{\rm
CMB})^2/(t_{\rm pix}N_{\rm pix})$ \cite{Kno95} is the variance per unit area
on the sky for polarization observations  when $t_{\rm pix}$ is the time spent on
each of $N_{\rm pix}$ pixels with detectors of NET $s$, and $\theta
\simeq 203\,f_{\rm sky}^{1/2}$ is roughly the width (in
degrees) of the survey.  In restricting the sum to
$l>180/\theta$, we have assumed that no information from modes
with wavelengths larger than the survey size can be obtained; in
fact, some information can be obtained, and our results should
thus be viewed as conservative \cite{LewChaTur02}.

\begin{figure}[t]
\centerline{\psfig{file=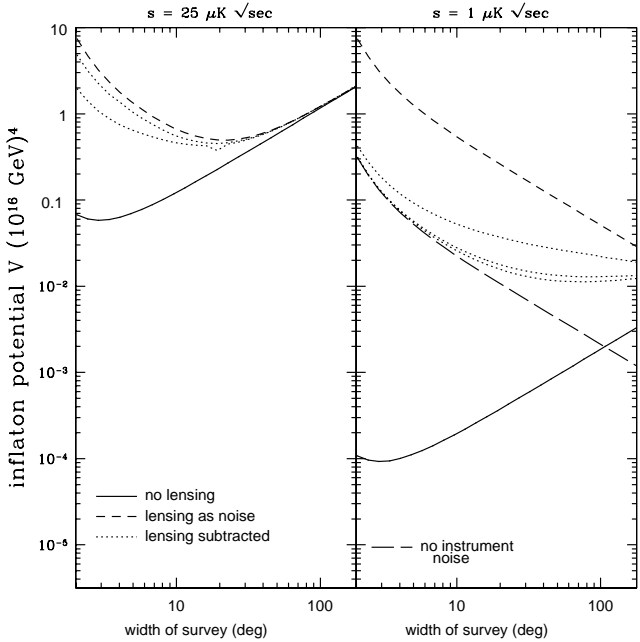,width=3.5in}}
\caption{Minimum inflation potential observable at
     $1\sigma$ as a function of survey width for a one-year
     experiment.  The left panel shows an experiment
     with NET $s=25\, \mu{\rm
     K}~\sqrt{\rm sec}$.
     The solid curve shows results assuming no CS
     while the dashed curve shows results including the effects
     of an unsubtracted CS; we take $\theta_{\rm
     FWHM}=5'$ in these two cases.  The dotted curves
     assume the CS is subtracted with $\theta_{\rm
     FWHM}=10'$ (upper curve) and $5'$ (lower curve).  Since
     the dotted curves are close to the dashed curve, it
     shows that these higher-order correlations will not be significantly
     useful in reconstructing the primordial curl for an
     experiment similar to Planck's sensitivity and resolution.  The
     right panel shows results for hypothetical improved
     experiments.  The dotted curves shows results with CS subtracted and 
     assuming $s=1\, \mu{\rm
     K}~\sqrt{\rm sec}$, $\theta_{\rm FWHM}=5'$, $2'$, and $1'$
     (from top to bottom). 
     The solid curve assumes $\theta_{\rm
     FWHM}=1'$ and $s=1\, \mu{\rm K}~\sqrt{\rm sec}$, and
     no CS, while the dashed curve treats CS as an additional
     noise.  The long-dash curve assumes  CS subtraction
     with no instrumental noise ($s=0$).}
\label{fig:infpot}
\end{figure}

\begin{figure}[t]
\centerline{\psfig{file=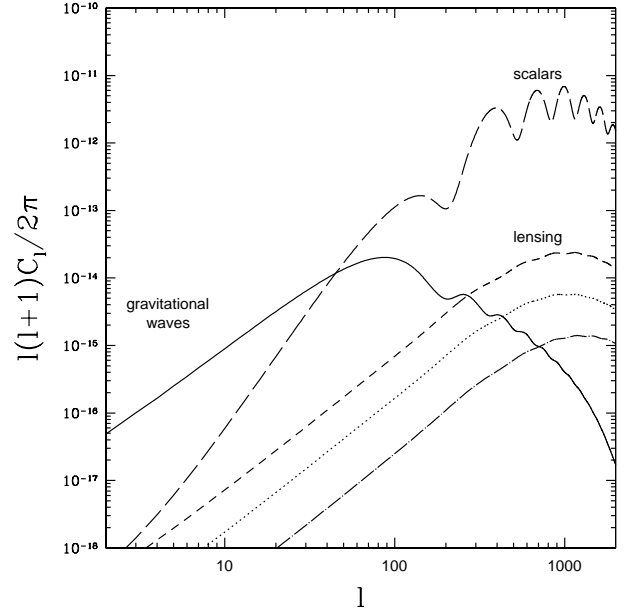,width=3.5in}}
\caption{
     Contributions to the CMB polarization power spectra.  The
     long-dashed curve shows the dominant polarization signal in the
     gradient component due to scalar perturbations. The solid
     line shows the maximum allowed curl polarization signal from the
     gravitational-wave background, which will be smaller if
     the inflationary energy scale is smaller than the maximum
     value allowed by COBE of $3.47 \times 10^{16}$ GeV.  
	The dashed curve shows the power
     spectrum of the curl component of the polarization due to
     CS.  The dotted curve is the CS contribution to the curl
     component that comes from structures out to a redshift of
     1; this is the level at which low-redshift lensing surveys
     can be used to separate the CS-induced polarization from
     the IGW signal. The dot-dashed line is the residual when
     lensing contribution is separated with a no-noise
     experiment and 80\% sky coverage.}
\label{fig:CBBlens}
\end{figure}

The second term in Eq. (\ref{eqn:sigma}) is due to
instrumental noise, and the first is due to cosmic variance.  In
the absence of CS, and for the null hypothesis of no IGWs, we
set $C_l^{BB}=0$, and the results for the smallest detectable
IGW amplitude are shown as the solid curves in
Fig. \ref{fig:infpot} for an experiment with detectors
of comparable sensitivity to Planck's (left) and a hypothetical experiment (right) with better
sensitivity. The smallest detectable IGW amplitude ${\cal T}$ scales as $s^2
t_{\rm yr}^{-1}$.  For large survey widths, it scales as
$\theta$, but at survey widths smaller than $\sim 5^\circ$ it
increases because information from the larger-angle modes in the
IGW-induced curl power spectrum is lost (cf. the IGW power
spectrum in Fig. \ref{fig:CBBlens}).

It is now easy to see how inclusion of CS affects
these results. As discussed above, lensing of the gradient polarization at the surface of
last scatter due to density perturbations leads to a CMB curl
component with a power spectrum,
\begin{equation} 
     \tilde{C}_{l}^{BB} = \frac{1}{2} \int \frac{d^2 \vecl_1}{(2 \pi)^2}
     \left[ \vecl_2 \cdot \vecl_1 \right]^2 (1 - \cos 4 \phi_{l_1})
     C_{l_2}^{\phi \phi} C_{l_1}^{EE} \, ,
\label{eqn:CBB}
\end{equation}
where $\vecl_2 = \vecl - \vecl_1$ here and throughout,
$C_l^{EE}$ is the power spectrum of the gradient component
of polarization and $C_l^{\phi\phi}$ is the power spectrum of
the projected lensing potential \cite{Hu00}. The latter is defined in terms
of the potential fluctuations, $\Phi$, along the line of sight
such that
\begin{eqnarray}
     \phi(\bn)&=&- 2 \int_0^{\rad_0} d\rad
     \frac{\da(\rad_0-\rad)}{\da(\rad)\da(\rad_0)}
     \Phi (\rad,\hat{{\bf n}}\rad ) \,,
\label{eqn:lenspotential}
\end{eqnarray}
where $\rad$ is the comoving radial distance, or conformal 
look-back time, with $r_0$ 
at the last scattering surface, and $\da(\rad)$ is the comoving angular diameter distance.  The
CS-induced curl power spectrum is shown as the dashed curve in
Fig. \ref{fig:CBBlens}.

By the time these measurements are made, the cosmological
parameters that determine this lensed curl power spectrum should
be sufficiently well determined that this power spectrum can be
predicted with some confidence.  In that case, the CS-induced
curl component can be treated simply as a well-understood noise
for the IGW background.  The smallest detectable IGW amplitude
can then be calculated as above, but now inserting the lensed
power spectrum, Eq. (\ref{eqn:CBB}), in Eq. (\ref{eqn:sigma}) for $C_l^{BB}$.
The results are shown as the short-dash curves in
Fig. \ref{fig:infpot}.  When lensing is included, the results
no longer scale simply with $f_{\rm sky}$, $s$, or $t_{\rm yr}$,
as there is now a trade-off between the instrumental-noise and
CS-noise terms in Eq. (\ref{eqn:sigma}).  The left panel shows
that the IGW sensitivity for an experiment with NET similar to
Planck's should not be affected by CS.  This is
because the IGW amplitudes that could be detectable by such experiments
are still relatively large compared with the expected CS signal,
especially at the larger angles that will be best accessed by
Planck. However, CS will affect the ability of experiments more
sensitive than Planck to detect unambiguously IGWs, as shown in
the right panel.  CS also
shifts the preferred survey region to larger areas, as the IGW
power spectrum peaks at larger angles than the CS power spectrum
(cf. Fig. \ref{fig:CBBlens}).  Finally, note that if the CS curl
is treated as an unsubtracted noise, it leads, assuming a
no-noise polarization map, to a smallest detectable IGW
amplitude, corresponding to an inflaton-potential height,
$V^{1/4} \sim 4 \times 10^{15}$ GeV.

Now we arrive at the main point of this {\it Letter}; i.e., how
well can the CS-induced curl be subtracted by mapping the CS as
a function of position on the sky?  One possibility is that the
primordial polarization pattern might be reconstructed from that 
observed by using CS maps obtained with correlations of galaxy
ellipticities \cite{wl}.  However, the source galaxies for these CS
surveys are at redshifts $z\sim1$, while only a small fraction
of the CS-induced curl comes from these redshifts, as indicated in
Fig. \ref{fig:CBBlens}.  An alternative possibility is to use
higher-order correlations in the CMB \cite{SelZal99}
to map the CS-induced curl all the way back to the surface of
last scatter.

CS modifies the temperature and polarization pattern, giving
rise to anisotropic correlations on
small scales where the image of the CMB surface of last scatter
is sheared by weak lensing.  According to Ref. \cite{Hu01}, the
quantity, ${\mathbf \nabla} \cdot [T(\bn) {\mathbf \nabla}
T(\bn)]$, provides the best quadratic estimator, given a temperature map,
of the deflection angle at position ${\mathbf \hat n}$ on
the sky.  In Fourier space, we can write this quadratic estimator
for the deflection  angle as
\begin{eqnarray}
     && \hat{\alpha}(\vecl) = \frac{N_l}{l} \int
     \frac{d^2\vecl_1}{(2\pi)^2} \left(\vecl 
     \cdot \vecl_1 C_{l_1}^{\rm CMB} +\vecl \cdot \vecl_2
     C_{l_2}^{\rm CMB}\right) \frac{T(l_1)T(l_2)}{2 C_{l_1}^{\rm
     tot} C_{l_2}^{\rm tot}} \, ,
\end{eqnarray}
where
$C_l^{\rm CMB}$ is the unlensed CMB power spectrum and
$C_l^{\rm tot}= C_l^{\rm lensed-CMB} + f_{\rm sky} w^{-1} e^{l^2 \sigma_b^2}$
includes all contributions to the CMB temperature power spectrum.  
The ensemble average over CMB realizations, $\langle
\hat{\alpha}(\vecl)_{\rm CMB} \rangle$, is equal to the deflection angle,
$l\phi(\vecl)$, when 
\begin{equation} 
\label{E:Nl}
     N_l^{-1} = \frac{1}{l^2}
     \int \frac{d^2\vecl_1}{(2\pi)^2} \frac{\left(\vecl \cdot
     \vecl_1 C_{l_1}^{\rm CMB} +\vecl \cdot \vecl_2
     C_{l_2}^{\rm CMB}\right)^2}{2 C_{l_1}^{\rm tot}
     C_{l_2}^{\rm tot}} \, .
\end{equation}
It can also be shown \cite{Hu01} that $N_l$ is the noise power
spectrum associated with the reconstructed deflection angle
power spectrum,
\begin{equation}
     \langle \hat{\alpha}(\vecl) \hat{\alpha}(\vecl') \rangle 
     = (2\pi)^2 \delta_D(\vecl+\vecl')
     \left(l^2 C_l^{\phi \phi} +N_l\right) \, .
\label{eqn:nl}
\end{equation}
Here the ensemble average is taken independently over realizations of both
the CMB and the intervening large-scale structure.
In addition to these temperature estimators for the deflection
angle, we also use analogous ones constructed from the
polarization, as discussed in Ref. \cite{HuOka01}, although we
do not reproduce those formulas here. The total noise in
the estimator for the deflection angle can then be constructed
by summing the inverses of the individual noise
contributions. We thus determine the variance with which 
each Fourier mode of $\phi$ can be reconstructed.

With the deflection angle obtained this way as a function of
position on the sky, the polarization at the CMB surface of last
scatter can be reconstructed (details to be presented elsewhere
\cite{KesCooKam02}).  In the ideal case, there would be no error
in the CS reconstruction leading to no residual
lensing-induced curl component.  Realistically, however, there
will be some error in the CS reconstruction, from measurement
error and also from cosmic variance, with a
noise power spectrum, $C_l^{\phi\phi,{\rm noise}}$, given by $N_l/l^2$.
Since $C_l^{\rm lensed-CMB} \gg C_l^{\rm CMB}$ along the damping tail of the CMB power spectrum,
$N_l$ converges to a finite value such that beyond a certain small angular scale, additional information from CMB
provides no further information on CS. This leads to a finite limit on lensing extraction and, subsequently, a limit
for the amplitude of IGW contribution that can be separated from CS.

The lensing reconstruction from CMB data only allows
the extraction of $C_l^{\phi\phi}$ to a multipole of
$\lesssim1000$ \cite{HuOka01}, but there is substantial
contribution to the CS-induced curl component from lensing at
smaller angular scales.  We thus replace $N_l/l^2$ by
$C_l^{\phi\phi}$ when the former exceeds the latter at large $l$.
This provides an estimate to the noise expected in the
reconstructed curl component that follows from implementing a
filtering scheme where high-frequency noise in the CS
reconstruction is removed to the level of the expected CS
signal.  The dot-dash curve in Fig. \ref{fig:CBBlens} shows the
residual CS-induced curl component that remains after subtraction.

We can now anticipate the smallest IGW amplitude detectable by
a CS-corrected polarization map by simply using this residual
noise power spectrum in Eq. (\ref{eqn:CBB}).  Results are shown as
dotted curves in Fig. \ref{fig:infpot}.  The left panel shows
results as a function of survey size for an experiment with NET
similar to Planck, while the right-hand panel shows results for
experiments with better sensitivity and resolution.
Since the dotted curves are just below the dashed curve in
the left-hand panel of Fig. \ref{fig:infpot}, we learn that Planck's
sensitivity will not be sufficient to warrant an
effort to reconstruct the primordial curl and  we would do
just as well to simply treat the CS-induced curl as a noise
component of known amplitude.
We can expect to improve the discovery reach for IGWs
by increasing the sensitivity and resolution.  The
right-hand panel of Fig. \ref{fig:infpot} shows results for a
hypothetical experiment with $s=1\, \mu{\rm K}~\sqrt{\rm
sec}$ and angular resolutions of $5'$, $2'$ and $1'$.  We now see
there is a significant difference between the dashed curve and
the dotted lines suggesting the increasing improvement with
increasing resolution.

Finally, to indicate the ultimate limits of this class of
experiments, the long-dash curve in Fig. \ref{fig:infpot} shows the
results assuming perfect detectors (i.e., $s=0$).  If there were
no CS-induced curl, then we would have sensitivity to an
arbitrarily small IGW amplitude, but the existence of a
CS-induced curl provides an ultimate limit of $V^{1/4}\simeq 4
\times 10^{15}$ GeV, as discussed above.  Correction for the
effects of CS with the CS map inferred from
higher-order correlations would allow us to access lower IGW
amplitudes. If there is no instrumental-noise limitation, the sensitivity to an IGW
signal is maximized by covering as much sky as possible, and the
lowest accessible inflaton potential, $\sim10^{15}$ GeV, is
obtained with a nearly all-sky experiment.

To conclude, we have studied the IGW amplitudes accessible by
mapping the curl component of the CMB polarization, taking into
account the effects of a CS-induced curl that is either modeled
as an unsubtracted noise or subtracted with a CS map obtained
with higher-order correlations.  We find that the CS
reconstruction is unlikely to improve the IGW discovery reach
of Planck.  To go beyond Planck, however, a CS map will need to
be constructed with temperature and polarization maps of higher sensitivity and
resolution than Planck.  An ultimate limit of roughly $V^{1/4}\sim10^{15}$
GeV to the detectable IGW amplitude using the techniques
considered here comes from the existence of finite CMB power on small angular scales.  
There are several possible ways this lower limit may
be improved upon.  First of all, we have used only the lowest-order
temperature-polarization correlations to reconstruct the CS.  The
inclusion of the complete temperature-polarization four-point correlation
functions and higher-order correlations may possibly improve the
CS reconstruction.
We have neglected reionization in our analysis.  Reionization
will boost the large-angle CMB polarization \cite{Zal97} improving the detectability of IGWs.
Another improvement in this limit may be achieved by relaxing our assumption
that power in the CMB drops exponentially at small scales.  If
the excess small-scale power recently detected by CBI \cite{CBI}
comes from high redshifts, then there will be more small-scale
coherence patches with which to reconstruct the CS.  In this
case, it is imaginable that a far more precise CS map can be
reconstructed, but this might require even better angular
resolution and sensitivity. 

\smallskip
During the preparation of the paper, we learned of other
very recently completed work by Knox and Song \cite{KnoSon02}
that performs a very similar calculation and reaches similar
conclusions.  This work was supported in part by NSF
AST-0096023, NASA NAG5-8506, and DoE DE-FG03-92-ER40701.  Kesden
acknowledges the support of an NSF Graduate Fellowship and AC
acknowledges support from the Sherman Fairchild Foundation.

\end{document}